\documentclass[twocolumn,preprintnumbers,amsmath,amssymb]{revtex4}
\usepackage{dcolumn}
\usepackage[dvips]{epsfig}
\usepackage{times}

\newcommand{\CKAF}{Cs$_{2}$KAgF$_{6}$}

\begin{document}       

\date{\today}

\title{\bf Rare case of magnetic Ag$^{3+}$ ion: double perovskite Cs$_{2}$KAgF$_{6}$}

\author{Ting Jia$^{1}$, Xiaoli Zhang$^{1}$, Ting Liu$^{1}$, Fengren Fan$^{2}$, Zhi
Zeng$^{1,3}$\footnote{Corresponding author.
zzeng@theory.issp.ac.cn}, X. G. Li$^{4}$, D. I. Khomskii$^{5}$, and Hua Wu$^{2}$\footnote{Corresponding author.
wuh@fudan.edu.cn}}

\affiliation{$^{1}$Key Laboratory of Materials Physics, Institute of
Solid State Physics, Chinese Academy of Sciences, Hefei 230031, China}

\affiliation{$^{2}$Laboratory for Computational Physical Sciences
(MOE), State Key Laboratory of Surface Physics, and Department of
Physics, Fudan University, Shanghai 200433, China}

\affiliation{$^{3}$Department of Physics, University of Science and
Technology of China, Hefei, 230026, China}

\affiliation{$^{4}$Hefei National Laboratory for Physical Sciences
at Microscale, Department of Physics, University of Science and
Technology of China, Hefei 230026, China}

 \affiliation{$^{5}$Department of Physics, University
of Cologne, Cologne 50937, Germany}

\date{\today}
\begin{abstract}

Normally $4d$ or $5d$ transition metals are in a low-spin state.
Here using first-principles calculations, we report on a rare case
of a high-spin $S$=1 magnetic state for the Ag$^{3+}$ ion in the
double perovskite Cs$_{2}$KAgF$_{6}$. We also explored a possibility
of a conventional low-spin $S$=0 ground state and find an associated tetragonal
distortion to be 0.29 {\AA}. However, the lattice elastic energy
cost and the Hund exchange loss exceed the \emph{e$_{g}$}
crystal-field energy gain, thus making the low-spin tetragonal
structure less favorable than the high-spin cubic structure. We
conclude that the compact perovskite structure of
Cs$_{2}$KAgF$_{6}$ is an important factor in stabilizing the unusual
high-spin ground state of Ag$^{3+}$.

\end{abstract}

\maketitle

\clearpage

\makeatletter
    \newcommand{\rmnum}[1]{\romannumeral #1}
    \newcommand{\Rmnum}[1]{\expandafter\@slowromancap\romannumeral #1@}
  \makeatother

\section{\bf Introduction}

Transition metal compounds possess diverse properties, largely due
to their electronic correlation effects and the interplay of charge, spin, and orbital
degrees of freedom. For many of them, spin state is an important
issue. It can be a high-spin (HS), low-spin (LS), or even an
intermediate-spin (IS) state. The corresponding electronic
configurations, e.g., for a Fe$^{3+}$ ion in an octahedral crystal
field, are $t_{2g}^3e_g^2$, $t_{2g}^5$, and $t_{2g}^4e_g^1$. The
spin state is intimately related to materials properties such as
magnetic transitions, metal-insulator transition, and transport
behavior. In the past decades, a lot of research has been carried
out on the spin state of 3\emph{d} transition-metal compounds
~\cite{Wang, Blaha1, Raccah, Zhang, pek, Jia, Rueff, Chang}.  A
consensus has been reached that the spin state issue arises mainly
from a competition between the Hund's rule coupling and crystal-field
splitting. The former favors a HS state, and the later a LS state.
Compared to the well studied spin state issue of 3\emph{d}
transition metals, little research has been done on the spin state
of 4\emph{d}/5\emph{d} transition-metal compounds\cite{Itoh, Wu3},
which is probably due to a fact that in most cases, 4$d$/$5d$
transition metals are in a LS state. As 4\emph{d}/5\emph{d} orbitals
are spatially more extended than 3\emph{d} ones, their stronger
interaction with the anionic ligands and a larger crystal-field
splitting but a moderate or weak Hund exchange normally stabilize
the LS state\cite{Itoh, Wu3}.

\begin{figure}[b]
\includegraphics[scale=0.6]{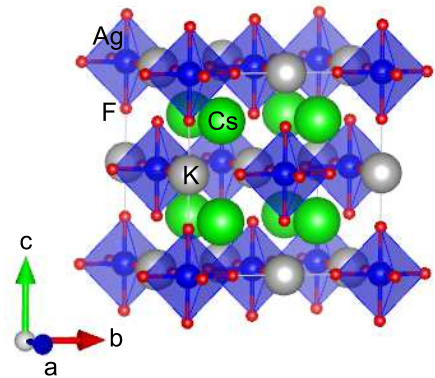}
\caption {(color online) Double perovskite structure of
Cs$_{2}$KAgF$_{6}$.}
\end{figure}

In this work, we have studied the double perovskite {\CKAF} using
first-principles calculations, focusing on its exotic spin state.
Fig. 1 shows its cubic structure, in which K$^{+}$ and Ag$^{3+}$
ions form an ordered arrangement on the $B$-sites of the perovskite
$AB$F$_3$~\cite{Hoppe,Grochala}, and the Ag-F bond distance is 2.13
\AA~ (see also the structural parameters in Table I). Although a
Ag$^{3+}$ ion ($4d^8$) could in principle be either in a LS state
($S$=0, $t_{2g}^6(3z^2-r^2)^2(x^2-y^2)^0$) or in a HS state ($S$=1,
$t_{2g}^6(3z^2-r^2)^1(x^2-y^2)^1$), in reality almost in all known
cases the Ag$^{3+}$ ion is in the LS state and is thus
nonmagnetic\cite{Mueller}. Actually, very few Ag$^{3+}$ and
Cu$^{3+}$ (both with $d^8$ electronic configuration) crystalline
compounds are known. These compounds mostly have a tetragonal or
lower symmetry structure to maintain the LS $S$=0 ground
state\cite{footnote}. Similar to the well-known Jahn-Teller (JT)
ions like Cu$^{2+}$ and Ag$^{2+}$, the Cu$^{3+}$ and Ag$^{3+}$ ions
could have an even stronger tendency to JT-driven distortions (local
elongation of ligand octahedra). Such distortions are usually so
strong that the corresponding energy gain exceeds the intraatomic
Hund's exchange energy, thus stabilizing the LS state. Indeed, the
energy of the state with the tetragonal distortion $u$ can be
written as
\begin{equation}
 E=-gnu+Bu^{2}/2,
 \end{equation}
where $g$ is the JT coupling constant determining the splitting of
the originally degenerate $e_g$ levels with the distortion, and $n$
is the number of electrons or holes on the $e_g$ levels (in our case
$n$ = 2, in contrast to $n$ = 1 (one hole) in the more familiar ion
Cu$^{2+}$ ($d^9$)). The second term on the right of the above
equation corresponds to lattice elastic energy. Minimizing this equation
with respect to the distortion $u$, we get the equilibrium
distortion
\begin{equation}
 u_{0}=gn/B,
 \end{equation}
so that the energy minimum is
\begin{equation}
 E_{0}=-g^{2}n^{2}/2B.
 \end{equation}
Thus we see that the decrease of the energy for the ions like
Cu$^{3+}$ and Ag$^{3+}$ with $n$ = 2 is {\it four times as large as} in the
already strong JT ions like Cu$^{2+}$. In effect this energy almost
always exceeds the on-site Hund's exchange energy  $J_H$, and such
ions have a very strong tendency to such distortion and to the
nonmagnetic LS state.

Although {\CKAF} was not much studied, the available magnetic
susceptibility and crystal structure data both suggest a HS state
for the Ag$^{3+}$ ion\cite{Grochala,Hoppe}. It is indeed very
surprising that {\CKAF} has magnetic Ag$^{3+}$ ions and these ions
sit in the undistorted F$_6$ octahedra. Why is this the case, which
physical mechanisms determine this and what is so specific for this
particular material, require explanation. Better understanding of
this could open possibilities to stabilize strange states like
$\texttt{"}$magnetic silver$\texttt{"}$ or $\texttt{"}$magnetic gold$\texttt{"}$ in some other situations.
Clarification of the physics leading to such unusual situation is
the main goal of the present investigation. Our first-principles
calculations indeed show that  the undistorted magnetic HS state of
Ag$^{3+}$ in {\CKAF} is the lowest energy state, and the more
conventional JT-distorted LS state lies much higher in energy. The
detailed analysis of our results demonstrates that the compact
perovskite structure of {\CKAF} could allow only a moderate
tetragonal distortion for the LS state,  due to a large elastic
energy cost associated with the $\texttt{"}$hard
lattice$\texttt{"}$. Then the moderate $e_g$ crystal-field energy
gain is not sufficient to compensate for the elastic energy cost and
the Hund's exchange loss. As a result, the LS state cannot be
stabilized as a ground state. Therefore, {\CKAF} is in the HS ground state
and represents a rare case of the magnetic Ag$^{3+}$ ion.

\begin{table}[tb]
\caption{The lattice parameters of Cs$_{2}$KAgF$_{6}$ in the experimental
cubic structure\cite{Hoppe}, and in the optimized tetragonal and ac-tetragonal phases.}

\begin{ruledtabular}
\begin{tabular}{cccc}

phases&cubic&tetragonal&ac-tetragonal\\
 \hline
space group&\emph{F}\emph{m}$\overline{3}$\emph{m}&\emph{I}4/\emph{m}\emph{m}\emph{m}&\emph{I}4/\emph{m}\emph{m}\emph{m}\\

a ({\AA})&9.175&6.488&6.195\\
b ({\AA})&9.175&6.488&6.195\\
c ({\AA})&9.175&9.175&10.062\\

x,y,z (Cs)&(0.75, 0.25, 0.25)&(0, 0.5, 0.25)&(0, 0.5, 0.25)\\
x,y,z (K)&(0.5, 0, 0)&(0, 0, 0)&(0, 0, 0)\\
x,y,z (Ag)&(0, 0, 0)&(0, 0, 0.5)&(0, 0, 0.5)\\
x,y,z (F)&(0, 0, 0.232)&(0.213, 0.213, 0.5)&(0.224, 0.224, 0.5)\\
&&(0, 0, 0.745)&(0, 0, 0.724)\\

\end{tabular}
\end{ruledtabular}
\end{table}


\section{\bf Computational Details}

We have used the experimental cubic structure of {\CKAF} and the
optimized tetragonal structures, see Table I. All the calculations
were performed using the full-potential augmented plane wave plus
local orbital method (Wien2k) \cite{Blaha}. The muffin-tin sphere
radii were chosen to be 2.8, 2.5, 2.2, and 1.4 bohr for Cs, K, Ag
and F atoms, respectively. The cutoff energy of 16 Ryd was used for
plane wave expansion. The calculations were fully converged using
300 k points in the first Brillouin zone. Exchange and correlation
effects were taken into account in a local spin density
approximation (LSDA)\cite{Perdew}. All the calculations were performed
in a ferromagnetic state unless otherwise
specified (Actually, {\CKAF} could well be paramagnetic, see the
results below). To account for the strong electron correlations, we
performed LSDA+\emph{U} calculations\cite{Anisimov}, where $U_{eff}$
=  $U$--$J$ ($U$ and $J$ are on-site Coulomb interaction and
exchange interaction, respectively) was used instead of
\emph{U}\cite{Dudarev}. And the orbital-dependent potential has the
form of $\Delta$\emph{V} = --$U_{eff}$(\emph{$\hat{n}$}$^{\sigma}$
-- $\frac{1}{2}$\emph{I})\cite{Madsen}, where
\emph{$\hat{n}$}$^{\sigma}$ is the orbital occupation matrix of spin
$\sigma$. This type of double-counting correction is made in a fully
localized limit\cite{Anisimov2,Laskowski}. In particular,
LSDA+\emph{U} method allows us to access different spin and orbital
states by initializing their corresponding density matrix. We find
that the calculated results change insignificantly for the $U_{eff}$
value in the range of 3-5 eV (which is a reasonable choice according
to previous studies on the electronic structure of
Cs$_2$AgF$_4$\cite{Rosner,Wu4}): The cubic HS state is constantly
more stable than the tetragonal LS state by more than 300 meV/fu
(see Table II for $U_{eff}$ = 4 eV), and the HS-LS relative energy
varies within only 10 meV/fu for $U_{eff}$ = 3-5 eV.
Correspondingly, with the increase of $U_{eff}$ from 3 to 5 eV, the insulating gap increases from 0.97 to 1.27 eV,
and the local Ag spin moment from 0.94 to 0.97 $\mu_B$. To further
confirm our LSDA+\emph{U} results, we also performed hybrid
functional PBE0 calculations~\cite{hyb,Adamo}, in which 1/4 Fock
exact exchange is mixed into the PBE exchange functional. Our PBE0
results show that the cubic HS state is more stable than the
tetragonal LS state by about 600 meV/fu, and the HS ground state has
an insulating gap of 1.48 eV and the local Ag spin moment of 1.06
$\mu_B$. All these results indeed (qualitatively) confirm the LSDA+$U$ results. As
such, below we focus on the results obtained with $U_{eff}$ = 4 eV.

\begin{table}[tb]
\caption{The relative total energies \emph{$\Delta$E} (meV/fu), band
gap \emph{E$_{g}$} (eV) and \emph{e$_{g}$} orbital occupation
(up/down) of Cs$_{2}$KAgF$_{6}$ in different states calculated by
LSDA+$U$. The \emph{t$_{2g}$} obitals are fully occupied in all
cases and omitted here. The cubic HS state is the ground state.}

\begin{ruledtabular}
\begin{tabular}{lrccc}

State&\emph{$\Delta$E}&\emph{E$_{g}$}&$3z^{2}-r^{2}$&$x^{2}-y^{2}$\\
 \hline
cubic HS&0&1.12&0.95/0.48&0.95/0.48\\
tetragonal LS&366&0.11&0.94/0.94&0.49/0.49\\
tetragonal HS&441&0.03&0.94/0.47&0.97/0.45\\
ac-tetragonal LS&336&0.23&0.94/0.94&0.49/0.49\\

\end{tabular}
\end{ruledtabular}
\end{table}

\begin{figure}

\includegraphics[scale=0.6]{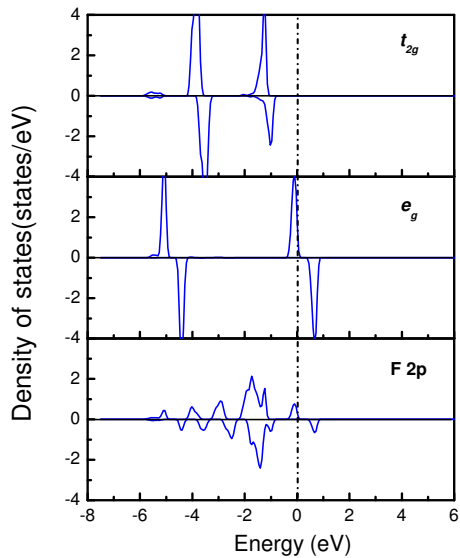}
\caption {(color online) The density of states of Ag-4\emph{d} and
F-2\emph{p} obitals for the cubic phase of Cs$_{2}$KAgF$_{6}$ by
LSDA.}

\end{figure}

\section{\bf Results and discussion}

To see first the nature of Ag-F interaction, a spin-polarized LSDA
calculation is performed for the cubic structure of
Cs$_{2}$KAgF$_{6}$ in a ferromagnetic state. We also calculated an
antiferromagnetic state and find that the exchange between two
nearest-neighboring Ag ions is weakly antiferromagnetic, being only
about 1 meV. The electronic structures of the ferromagnetic and antiferromagnetic
states are practically the same.
Considering a magnetic frustration associated with the
FCC sublattice of the Ag ions, a paramagnetic state, or a spin
glassy one below very low temperature (few K) can be expected for
Cs$_{2}$KAgF$_{6}$. Our LSDA calculations show that it is an
insulator with a band gap of 0.5 eV given by the $e_g$ exchange
splitting, see Fig. 2. Besides the common $t_{2g}$-$e_g$
crystal-field splitting, there is a large bonding-antibonding
splitting due to the strong Ag $4d$-F $2p$ covalency, being about 5
and 3 eV for the $e_g$ and $t_{2g}$ orbital, respectively. As a
result, the $e_g$ bonding state is even lower than the $t_{2g}$
one\cite{Ushakov}. The $t_{2g}$ bands are fully occupied and the
$e_g$ bands are 3/4 occupied, with the fully occupied up-spin
$t_{2g}$ and $e_g$ channels. This demonstrates the existence of a HS
state and a $d^9\underline{L}$ configuration for the formal
Ag$^{3+}$ ($d^8$) ion.  The one hole spreads over the six F atoms in
the AgF$_6$ unit, which accounts for a reduced spin moment of 0.90
$\mu_B$ on the Ag site and a finite spin moment of 0.16 $\mu_B$ on
each F. Summing up of all these moments and a small fraction in the
interstitial region gives a total integer spin moment of 2
$\mu_B$/fu, confirming the formal HS $S$=1 state for the Ag$^{3+}$
ion. As these magnetic AgF$_6$ units are separated in this double
perovskite by the corner-shared KF$_6$ units, they form narrow bands
 (sharp DOS curves) and are weakly antiferromagnetically
coupled.

Late $4d$ transition-metal oxides or fluorides often have a moderate
electron correlation. As {\CKAF} is a narrow band system, it is
natural to study its electron correlation effect. In the following,
we carried out LSDA+$U$ calculations to study the electron structure
of {\CKAF} and its spin state. By a comparison of Figs. 3 and 2,  we
see that with inclusion of the Hubbard's interaction \emph{U} the Ag
$4d$ states undergo some changes, and the biggest effect is that the
band gap is increased from 0.5 eV by LSDA to 1.1 eV by LSDA+$U$. The
HS $S$=1 state remains, and the strong Ag-F covalency is reflected
again by the occupied bonding state of the otherwise empty down-spin
$e_g$ orbital. It is the covalent electron occupation of
0.48$e$$\times$2 (see Table II) that accounts for the actual
configuration $d^9\underline{L}$ of the formal Ag$^{3+}$ ion which
is a negative charge transfer cation\cite{footnote}.

The above HS magnetic state of the formal Ag$^{3+}$ ion seems to be a straightforward result, but it is actually quite unusual as $4d$ transition metals are normally
in a LS state and the Ag$^{3+}$ is indeed in the LS state for almost
all known cases\cite{Mueller}. We are thus motivated to check
whether there exists another stable structural phase for {\CKAF} in
which the normal LS state could become a ground state. For this purpose,
we have also studied a tetragonal structure and carried out a structural optimization.
A tetragonal distortion would yield an $e_g$ crystal-field
splitting, and if the splitting is big enough, the LS state could
become the ground state.

\begin{figure}
\includegraphics[scale=0.6]{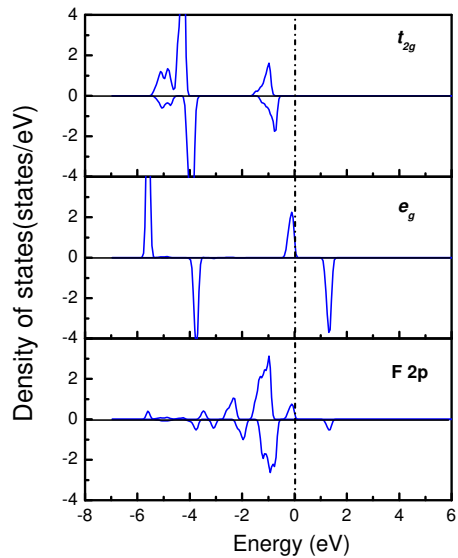}
\caption {(color online) The density of states of Ag-4\emph{d} and
F-2\emph{p} obitals for the cubic phase of Cs$_{2}$KAgF$_{6}$ by
LSDA+U.}
\end{figure}

To explore the possibility of the LS ground state, we lower the
lattice symmetry to allow a local tetragonal distortion of the
AgF$_6$ octahedra. We first use the above cubic lattice constants
but optimize the internal atomic positions under a tetragonal
symmetry, by minimizing the atomic forces\cite{Tran} till each one being
smaller than 25 meV/\AA. After a full electronic and atomic
relaxation associated with the initialized LS setup
$t_{2g}^6$($3z^2-r^2$)$^2$($x^2-y^2$)$^0$ for the formal Ag$^{3+}$
ion, the LS solution is indeed achieved. It is a local minimum of
the energy and has an elongated apical Ag-F bondlength of 2.25
\AA~ and a shrinking planar Ag-F bondlength of 1.96 \AA~ (see Table
I for the tetragonal phase). Thus, the LS solution is stabilized by
the tetragonal distortion (0.29 \AA, 2.25/1.96 = 1.15).
Moreover, in this tetragonally
distorted structure, the corresponding HS solution turns out to be
less stable than the LS state by 441--366 = 75 meV/fu, see Table II.
However, the tetragonal LS phase is 366 meV/fu higher in energy than
the above cubic HS one. We further calculated a regular tetragonal
structure with $c_0/a_0$ [i.e., $c/(\sqrt{2}a)$] = 1.15, see Table I
for the ac-tetragonal structure which preserves the volume of the
experimental cubic cell and the above tetragonal distortion. This
ac-tetragonal LS phase turns out to be still higher in energy than
the cubic HS phase by 336 meV/fu.

Now we carry out a volume optimization for both the cubic HS state
and the ac-tetragonal LS state. The results are shown in Fig. 4. The
equilibrium volume of the cubic HS state is about 11\% smaller than
the experimental one (i.e., within 4\% underestimation of the
lattice constant, within a few percent error bar of
density-functional calculations). The equilibrium volume of the
ac-tetragonal LS state is also 5\% smaller than the experimental
one. Note that the energy difference between the two equilibrium
states, the cubic HS and the ac-tetragonal LS states, increases up
to 750 meV/fu. This indicates that the cubic HS ground state is
robust against the structural optimization. The metastable LS state
is stabilized by the tetragonal distortion, which could be realized
in a biaxial pressure experiment. The present bistable HS and LS
solutions can also be understood by a pseudo Jahn-Teller mechanism
for an $e_g^2$ system\cite{Garcia11}.

\begin{figure}
\includegraphics[scale=0.6]{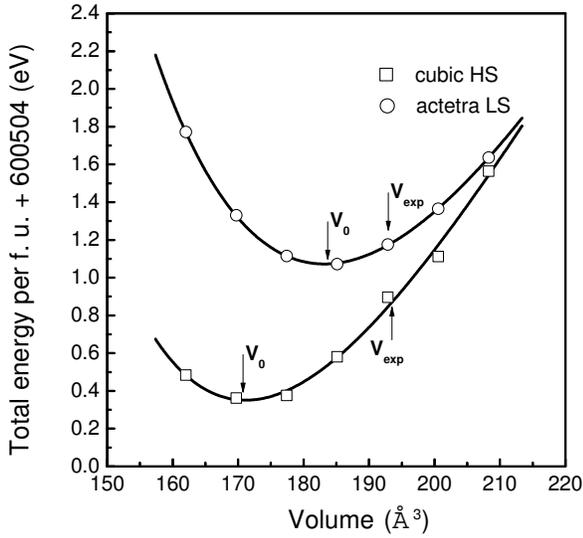}
\caption {Volume optimization of the cubic HS phase and the ac-tetragonal LS phase by LSDA+U.}
\end{figure}

\begin{figure}
\includegraphics[scale=0.6]{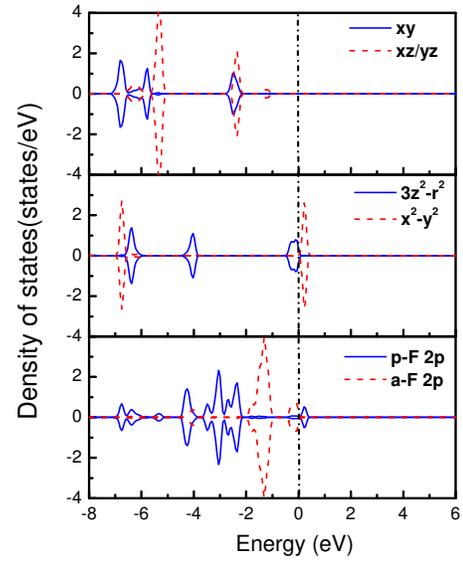}
\caption {(color online) The density of states of Ag-4\emph{d} and
planar/apical F-2\emph{p} obitals for the tetragonal LS phase of
Cs$_{2}$KAgF$_{6}$ by LSDA+U.}
\end{figure}

Before analyzing the reason why the cubic HS phase is the ground
state, we now have a close look at the tetragonal phase either in
the LS or HS state. In the tetragonal LS state, the degeneracies of
both \emph{e$_{g}$} and \emph{t$_{2g}$} states are lifted, see Fig.
5. The threefold degenerate \emph{t$_{2g}$} states are split into
doubly degenerate (\emph{d$_{xz}$}/\emph{d$_{yz}$}) states and
singlet \emph{d$_{xy}$} state. Similarly, the twofold degenerate
\emph{e$_{g}$} states split into nondegenerate
\emph{d$_{x^{2}-y^{2}}$} and \emph{d$_{ 3z^{2}-r^{2}}$} states. It
is also an insulator with a small band gap of 0.1 eV, which lies
between the occupied \emph{pd}$\sigma$ antibonding band of the
\emph{d$_{ 3z^{2}-r^{2}}$} orbital and the unoccupied
\emph{pd}$\sigma$ antibonding band of the \emph{d$_{x^{2}-y^{2}}$}
orbital. The equal occupation of the spin-up and spin-down orbitals
results in a nonmagnetic state without any exchange splitting, in
accordance with the LS $S$=0 character. Note that the strong
hybridization between Ag 4\emph{d} and planar-F/apical-F 2\emph{p}
orbitals is similar to that in cubic phase (see Table II), by which the Ag
4\emph{d} orbitals split into a series of narrow bonding and
antibonding orbitals. Such electronic structure, with narrow-band
and large bonding-antibonding splitting, is governed by the behavior of the isolated
AgF$_{6}$ clusters. Generally speaking, the bonding-antibonding
splitting of both \emph{e$_{g}$} orbitals is larger than that of
\emph{t$_{2g}$} orbitals due to larger overlap between
\emph{e$_{g}$} and \emph{p} orbitals. Furthermore, the
bonding-antibonding splitting of \emph{d$_{x^{2}-y^{2}}$} orbital in the tetragonal phase is larger than that of \emph{d$_{ 3z^{2}-r^{2}}$} due to the shrinking
planar Ag-F bond length. The \emph{t$_{2g}$} and \emph{d$_{
3z^{2}-r^{2}}$} bands are fully occupied and the
\emph{d$_{x^{2}-y^{2}}$} band is half occupied. Hence, the 4\emph{d}
occupation of 9$e$  is one more electron than the nominal value for
Ag$^{3+}$. This suggests the strong covalent nature of the Ag-F
bonds, in contrast to the naive ionic picture.

\begin{figure}
\includegraphics[scale=0.6]{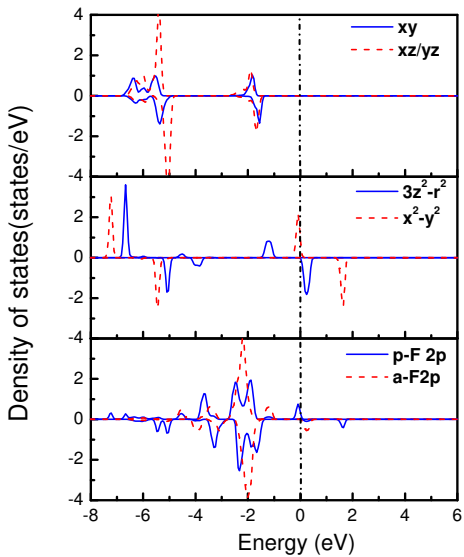}
\caption {(color online) The density of states of Ag-4\emph{d} and
planar/apical F-2\emph{p} obitals for the tetragonal HS phase of
Cs$_{2}$KAgF$_{6}$ by LSDA+U.}
\end{figure}

For the spin-polarized HS state of the tetragonal phase (Fig. 6),
the band splitting and Ag-F hybridization are similar to those in
the LS state, but the occupation of \emph{e$_{g}$} orbitals has a
remarkable change. While the up-spin \emph{d$_{x^{2}-y^{2}}$} and
\emph{d$_{ 3z^{2}-r^{2}}$} states are fully occupied, their
down-spin counterparts are half occupied. The unequal occupation of
the up-spin (5$e$) and down-spin (4$e$) states results in a considerable exchange
splitting, reflecting a HS state character. Although the double
degeneracy of \emph{e$_{g}$} orbitals is lifted by the tetragonal
distortion, the electron configuration of
(\emph{d$_{x^{2}-y^{2}}$})$^{0.5}$ and (\emph{d$_{
3z^{2}-r^{2}}$})$^{0.5}$ for down-spin is the same as that in cubic
phase with $e_g$ degeneracy, indicating again the significant Ag-F
covalency. A tiny band gap lies between the occupied
\emph{pd}$\sigma$ antibonding band of the \emph{d$_{ x^{2}-y^{2}}$}
orbital and the unoccupied \emph{pd}$\sigma$ antibonding band of the
\emph{d$_{ 3z^{2}-r^{2}}$} orbital.

Here we clarify the reason why the cubic HS phase is the lowest
energy state, using the above total-energy results (see Table II).
The 441 meV/fu increase of energy from the cubic HS phase to the
tetragonal HS phase is attributed to the lattice elastic energy cost
(\emph{E$_{el}$}) associated with the tetragonal distortion, as the electronic configuration
remains unchanged in both HS states and no crystal-field energy is involved. The
energy difference of 75 meV/fu between the tetragonal HS and
tetragonal LS (both in the same lattice) is due to the competition between the Hund exchange
($J_H$) and the crystal-field level splitting ($E_{CF}$) of the
$e_g$ electrons, with $E_{CF}$ and $J_H$ being both about 1 eV and
$E_{CF}$ slightly larger than $J_H$. Whether the cubic HS phase or
the tetragonal LS one is the ground state depends on three competing
contributions: $\Delta$\emph{E} = \emph{E}(tetragonal LS) --
\emph{E}(cubic HS) = \emph{E$_{el}$} + \emph{J$_{H}$} --
\emph{E$_{CF}$}. Taking  \emph{J$_{H}$} -- \emph{E$_{CF}$} = --75
meV/fu from above, the tetragonal LS phase could become the ground
state ($\Delta$\emph{E} $<$ 0) only when the \emph{E$_{el}$} is
lower than 75 meV/fu. However, here \emph{E$_{el}$} is up to 441 meV/fu.
Therefore, the cubic HS phase is the ground state, and the large
elastic energy cost in this compact double perovskite structure
prevents the tetragonal distortion from yielding a large $E_{CF}$ to
stabilize the LS state.

As seen in Figs. 5 and 6, in the tetragonal phase, the
\emph{e$_{g}$} crystal-field splitting and the exchange splitting
are quite close in size. This explains the small band gap present in
both the LS and HS solutions. This also accords with the small
energy difference of 75 meV/fu between the tetragonal HS and LS
phases. As discussed above, three competing contributions get
involved in determination of the ground spin-state solution:
$\Delta$\emph{E} = \emph{E}(tetragonal LS) -- \emph{E}(cubic HS) =
\emph{E$_{el}$} + \emph{J$_{H}$} -- \emph{E$_{CF}$}. Among them, the
lattice elastic energy cost $E_{el}$ and the \emph{e$_{g}$}
crystal-field splitting $E_{CF}$ are most influenced by the
structural details, but the Hund exchange $J_{H}$ is almost a
constant, being about 1 eV.  Therefore, only a large $E_{CF}$ but a
small $E_{el}$ would favor a tetragonal LS ground state. We thus
conceive that such a situation could be met in an open structure,
which can maintain a large Jahn-Teller distortion (and thus a large
$E_{CF}$) but with a low $E_{el}$ cost. For example, in the
well-known La$_{2}$CuO$_{4}$, there is a huge distortion (0.5 \AA)
of the Cu-O bonds, being 1.90 {\AA} $\times$ 4 in plane and 2.40
{\AA} $\times$ 2 out of plane\cite{Longo}. This is due to a lattice
strain effect in this layered material\cite{Wu09} and to the Cu-O
bondlength adjustment for the $x^2-y^2$ type hole-orbital order
driven by an anisotropic crystal field\cite{Wu11,Garcia14}. Upon a
hole doping, the holes spread over the in-plane ligand oxygens and
they hybridize with the central Cu$^{2+}$ $S$=1/2 to form a
Zhang-Rice singlet\cite{Zhang2} (the \emph{d$^{9}\underline{L}$}
configuration with \emph{S} = 0 or a nominal LS Cu$^{3+}$ with two
holes on \emph{d$_{x^{2}-y^{2}}$} orbital). It is the large $e_g$
crystal-field splitting but a small elastic energy cost which lead
to the formation of the LS state upon hole doping. We can expect to
have a similar situation in the layered 214 structure with Ag
instead of Cu. However, in the compact perovskite structure of
{\CKAF}, the tetragonal distortion cannot be as large as in the
layered La$_2$CuO$_4$. As a result, here the LS state is disfavored,
but the magnetic HS state of the Ag$^{3+}$ ion is the ground state.

\section{\bf Conclusions }

To summarize, using first-principles calculations, we find that the
cubic double perovskite {\CKAF} has a strong Ag-F covalency, and
that the formal Ag$^{3+}$ ion has an unusual high-spin (HS) $S$ = 1
ground state. This is in sharp contrast to a common view that
Ag$^{3+}$ always prefers a tetragonal or lower-symmetry coordination and a
low-spin (LS) $S$=0 state.
In calculations, we could also get the $\texttt{"}$normal$\texttt{"}$
LS state, associated with a tetragonal distortion of 0.29 \AA.
However, the tetragonal LS state turns out to be energetically
less favorable than the cubic HS state. We propose that three
competing contributions tip the balance between the HS and LS
states, which is formulated by the expression
$\Delta{E}=E_{LS}-E_{HS}=E_{el}+J_H-E_{CF}$. As the compact
perovskite structure of {\CKAF} allows only a moderate tetragonal
distortion (but with a large elastic energy cost $E_{el}$), a
corresponding moderate crystal-field energy gain $E_{CF}$ in the
tetragonal LS state cannot compensate for the $E_{el}$ and
Hund's exchange loss $J_H$. As a result, the cubic HS phase
is the ground state. We thus conclude that {\CKAF} is a rare case of
the magnetic Ag$^{3+}$ ion, and suggest that the Ag$^{3+}$ LS
nonmagnetic state can readily be formed in open structure materials
such as layered systems.

\section{\bf Acknowledgments}

This work was supported by the NSF of China (Grant No. 11204309,
U1230202), Anhui Province (Grant No. 1308085QA04), the special Funds
for Major State Basic Research Project of China (973) under Grant
No. 2012CB933702, Hefei Center for Physical Science and Technology
under Grant No. 2012FXZY004, and the Director Grants of Hefei
Institutes of Physical Science, Chinese Academy of Science (CAS).
The calculations were performed in Center for Computational Science
of CASHIPS and on the ScGrid of Supercomputing Center, Computer
Network Information Center of CAS. H. Wu is supported by the NSF of
China (Grant No. 11274070), Pujiang Program of Shanghai (Grant No.
12PJ1401000), and ShuGuang Program of Shanghai (Grant No. 12SG06).
D. Kh. is supported by the German Project FOR 1346 and by Cologne
University via German Excellence Initiative.

\end{document}